\documentclass[lettersize,journal]{IEEEtran}
\usepackage{amsmath,amsfonts}
\usepackage{algorithmic}
\usepackage[linesnumbered,ruled,vlined]{algorithm2e}
\usepackage{amsmath}
\usepackage{array}
\usepackage[caption=false,font=normalsize,labelfont=sf,textfont=sf]{subfig}
\usepackage{textcomp}
\usepackage{stfloats}
\usepackage{url}
\usepackage{verbatim}
\usepackage{graphicx}
\usepackage{cite}
\begin{document}

\title{Decision Transformer for IRS-Assisted Systems with Diffusion-Driven Generative Channels}

\author{Jie Zhang, Jun Li, Zhe Wang, Yu Han, Long Shi, and Bin Cao 
        \IEEEcompsocitemizethanks{
        \IEEEcompsocthanksitem Jie Zhang, and Long Shi are with the School of Electronic and Optical Engineering, Nanjing University of Science and Technology, Nanjing 210094, China (e-mail: zhangjie666@njust.edu.cn,  slong1007@gmail.com). 
        \IEEEcompsocthanksitem Jun Li is with the School of Information Science and Engineering,  Southeast University, Nanjing 210096, China (e-mail: jun.li@seu.edu.cn) ($\emph{Corresponding author}$)
        \IEEEcompsocthanksitem Zhe Wang is with the School of Computer Science and Engineering, Nanjing University of Science and Technology, Nanjing 210094, China (email: zwang@njust.edu.cn).
        \IEEEcompsocthanksitem Yu Han is with the National Mobile Communications Research Laboratory, Southeast University, Nanjing 210096, China (e-mail: hanyu@seu.edu.cn).
        \IEEEcompsocthanksitem Bin Cao is with the State Key Laboratory of Networking and Switching Technology, Beijing University of Posts and Telecommunications, Beijing 100876, China (e-mail: caobin65@163.com). 
        } 
        }

\maketitle

\begin{abstract}
In this paper, we propose a novel diffusion-decision transformer (D2T) architecture to optimize the beamforming strategies for intelligent reflecting surface (IRS)-assisted multiple-input single-output (MISO) communication systems. The first challenge lies in the expensive computation cost to recover the real-time channel state information (CSI) from the received pilot signals, which usually requires prior knowledge of the channel distributions. To reduce the channel estimation complexity, we adopt a diffusion model to automatically learn the mapping between the received pilot signals and channel matrices in a model-free manner. The second challenge is that, the traditional optimization or reinforcement learning (RL) algorithms cannot guarantee the optimality of the beamforming policies once the channel distribution changes, and it is costly to resolve the optimized strategies. To enhance the generality of the decision models over varying channel distributions, we propose an offline pre-training and online fine-tuning decision transformer (DT) framework, wherein we first pre-train the DT offline with the data samples collected by the RL algorithms under diverse channel distributions, and then fine-tune the DT online with few-shot samples under a new channel distribution for a generalization purpose. Simulation results demonstrate that, compared with retraining RL algorithms, the proposed D2T algorithm boosts the convergence speed by $3$ times with only a few samples from the new channel distribution while enhancing the average user data rate by $6\%$.
\end{abstract}


\section{Introduction}
Intelligent reflecting surfaces (IRS) have emerged as a significant innovation for wireless communications~\cite{IRS_background}. An IRS consists of multiple passive reflecting elements and a control circuit, where each element can independently alter the amplitude and phase of incident electromagnetic waves to adjust the signal's propagation path between the base station (BS) and users. The IRS can significantly improve the communication quality by optimizing the passive beamforming strategy according to the time-varying channel state information (CSI).  

In IRS-assisted communication systems, the IRS beamforming can be optimized through the conventional convex or heuristic optimization methods. The authors in \cite{IRS_FPI} have employed fixed point iteration strategies to jointly optimize the beamforming at the base station and that of the IRS, thereby maximizing the spectral efficiency. However, the traditional optimization methods often rely on precise modeling of the CSI distribution and suffer from high computational complexity.
In contrast, model-free methods based on reinforcement learning (RL) have demonstrated superior performance in dynamic and unknown IRS environments. The authors in \cite{IRS_DQN} and\cite{IRS_JieZhang} utilize the deep Q-network (DQN) algorithm of \cite{UAV_liu_qian} to optimize the phase shifts of IRS through trial-and-error manner without the need for complex mathematical modeling. Similarly, the authors in \cite{IRS_DDPG} utilize the deep deterministic policy gradient (DDPG) algorithm to optimize the beamforming between the BS and IRS in a continuous optimization space, effectively reducing the transmit power of the BS.

Despite the considerable potential of the RL methods, they still face challenges in adapting to the variations of communication environments. For example, the environment in which RL algorithms are pre-trained may not align with the environment in which policies are executed. In IRS-assisted communication scenarios, the distributions of CSI may change due to alterations in physical environments or the interfering sources. Therefore, pre-trained RL algorithms fail to generalize the beamforming policy effectively into new environments, which necessitates a complete retraining from the ground up with extensive data samples, and thereby consuming substantial training resources. Therefore, it is challenging to generalize the optimized beamforming policies for new channel distribution environments with rapid convergence.

To address the inefficiencies of RL in adapting to new environments, the introduction of decision transformers (DT)~\cite{decision_transformer} represents a significant advancement. Unlike traditional RL algorithms, DT employs a unique sequence modeling approach that leverages the powerful representation and generalization capabilities of the transformer architecture to effectively handle complex decision-making problems. In this way, DT can learn the mapping from historical sequences to optimal beamforming, which excels in generalizing learned policies to new scenarios without the need of training from scratch. Moreover, the effectiveness of the DT relies on precise channel estimations. However, traditional methods require significant computational resources due to the repeated calculations of mean values and covariance matrices of the prior channel model~\cite{Traditional_CE_issuses,Traditional_ce_issue_2}. To bridge this gap, a diffusion model (DM)~\cite{DDPM} can be introduced to obtain the real-time channel through supervised learning.


Driven by these issues, we present a novel DM-DT (D2T) structure that incorporates the strengths of DM and DT in IRS-assisted communication systems for the beamforming optimization. The main contributions are summarized as follows.
\begin{itemize}
    \item The proposed D2T structure utilizes the DM to recover the real-time channel from the received pilot signals. This channel acquisition method does not rely on the mathematical models required by traditional estimation methods and can approximate the actual channel closely, thus enhancing the efficiency and effectiveness of channel estimation in diverse distribution conditions.

    \item The proposed D2T algorithm collects datasets that include optimal beamforming strategies, CSI, and user throughput, which are gathered after RL algorithm converges across various channel distribution environments. These datasets are used to pre-train the DT model. In new channel distribution environment, this pre-trained decision model can achieve strategy generalization through fine-tuning with a small number of samples, thereby eliminating the need for retraining from scratch.

    \item Simulation results demonstrate that the pre-trained D2T method performs well in new channel distribution environment even with zero-shot learning. Our fine-tuned model converges with few-shot samples and outperforms the traditional RL methods by $6\%$ with a convergence speed that is $3$ times faster. Furthermore, the performance of the D2T method approaches the upper bounds achieved by the DT with perfect CSI.
\end{itemize}
$\emph{Notations:}$ Matrices and vectors are denoted by bold capital and lower-case letters, respectively. $\boldsymbol{A}^{ij}$ denotes the element in the $i$-th row and $j$-th column of the matrix $\boldsymbol{A}$. $(\cdot)^{H}$ and $(\cdot)^{T}$ denote the conjugate transpose and transpose operations, respectively. $\mathrm{diag}(\boldsymbol{a})$ is a diagonal matrix with entries of a vector $\boldsymbol{a}$ in main diagonal. $\mathbb{C}^{m\times n}$ and $\mathbb{R}^{m\times n}$ denote the sets of $m\times n$ matrices in complex domain and in real domain, respectively. $\emptyset$ denotes the empty set.

\section{System Model}
We consider a multiple-input single-output (MISO) system assisted by an IRS, where the BS is equipped with $M$ antennas, the IRS is composed of $N$ elements, and the user has a single antenna. Let us consider a quasi-static frequency-flat fading channel with a time-slotted system, and each time slot has a duration of $\tau$. Suppose that the direct link between the BS and the user is blocked. The channel coefficient matrix from the BS to the IRS at time slot $t$ is denoted as $\boldsymbol{\mathrm{G}}_{t}\in\mathbb{C}^{N\times M}$, and the channel coefficient vector from the IRS to the user at time slot $t$ is denoted as $\boldsymbol{\mathrm{h}}_{t}\in\mathbb{C}^{N\times 1}$. Hence, the received signal at the user can be expressed as
\begin{equation}\label{received signal 1}
    y_{t} = (\boldsymbol{\mathrm{h}}_{t}^{H}\boldsymbol{\Phi}_{t}\boldsymbol{\mathrm{G}}_{t})\boldsymbol{\mathrm{f}}_{t}s_{t} + n_{t},
\end{equation}
where $\boldsymbol{\Phi}_{t}=\mathrm{diag}(\boldsymbol{\phi}_{t})$ is the beamforming matrix and $\boldsymbol{\phi}_{t}=[\beta_{t,1}e^{j\phi_{t,1}},\dots,\beta_{t,N}e^{j\phi_{t,N}}]^{T}$ is the beamforming vector. $\beta_{t,n}\in\left[0,1\right]$ is the amplitude reflecting coefficient and $\phi_{t,n}\in\left[0,2\pi\right)$ is the phase reflecting coefficient for the $n$-th IRS element. $\boldsymbol{\mathrm{f}}_{t}\in\mathbb{C}^{M\times 1}$ is the beamforming vector at the BS with the power constraint $||\boldsymbol{\mathrm{f}}_{t}||^{2}\leq P$. $s_{t}$ is the data transmitted by the BS satisfying $\mathbb{E}\left[|s_{t}|^{2}\right]=1$. The additive white Gaussian noise with zero mean and $\sigma^{2}$ variance is denoted as $n_{t}$. Given~(\ref{received signal 1}), the received signal can be rewritten as 
\begin{equation}
    y_{t} = (\boldsymbol{\phi}_{t}^{T}\mathrm{diag}(\boldsymbol{\mathrm{h}}_{t}^{H})\boldsymbol{\mathrm{G}}_{t})\boldsymbol{\mathrm{f}}_{t}s_{t} + n_{t},
\end{equation}
where $\boldsymbol{\mathrm{H}}_{t}=\mathrm{diag}(\boldsymbol{\mathrm{h}}_{t}^{H})\boldsymbol{\mathrm{G}}_{t}$ is the cascaded channel between BS and user through IRS. For simplicity, we utilize the maximum-ratio transmission as the optimal IRS beamforming policy for a fixed phase shift vector $\boldsymbol{\phi}_{t}$~\cite{Max_Ratio_Trans}, i.e.,
\begin{equation}
    \boldsymbol{\mathrm{f}}^{*}=\sqrt{P}(\boldsymbol{\phi}_{t}^{T}\boldsymbol{\mathrm{H}_{t}})^{H}/||\boldsymbol{\phi}_{t}^{T}\boldsymbol{\mathrm{H}}_{t}||.
\end{equation}

Our objective is to optimize the IRS beamforming matrix $\boldsymbol{\Phi}_{t}$ to maximize the long term cumulative data rate, i.e.,
\begin{equation}\label{optimization function}
\begin{aligned}
\textbf{P}:&\max_{\left\{\boldsymbol{\Phi}_{t}\right\}}\sum_{t=1}^{T}\log_{2}\left(1+\frac{P||\boldsymbol{\phi}_{t}^{T}\boldsymbol{\mathrm{H}_{t}}||^{2}}{\sigma^{2}}\right)\\
    &\text{s.t.}\;\qquad |\boldsymbol{\Phi}_{t}^{nn}|=1, \forall n\in\left[1,\dots,N\right].
\end{aligned}
\end{equation}

\section{Proposed D2T Solution}
It is worth noting that there exist solutions to the optimization problem $\textbf{P}$ itself, such as the semi-definite relaxation (SDR) method and RL approaches~\cite{Jie_Zhang_IRS_2}. These methods assume a static channel distribution environment. However, when the CSI distribution varies, these approaches necessitate either restarting the iterative process or retraining the models from scratch. Such requirements significantly increase computational costs and hinder the generalization to new channel distribution environments. In this section, we first introduce preliminary concepts of DM and DT, and then propose a novel method to handle new IRS environments with rapid convergence.

\begin{figure*}[htbp]
    \centering
    \includegraphics[width=0.95\textwidth]{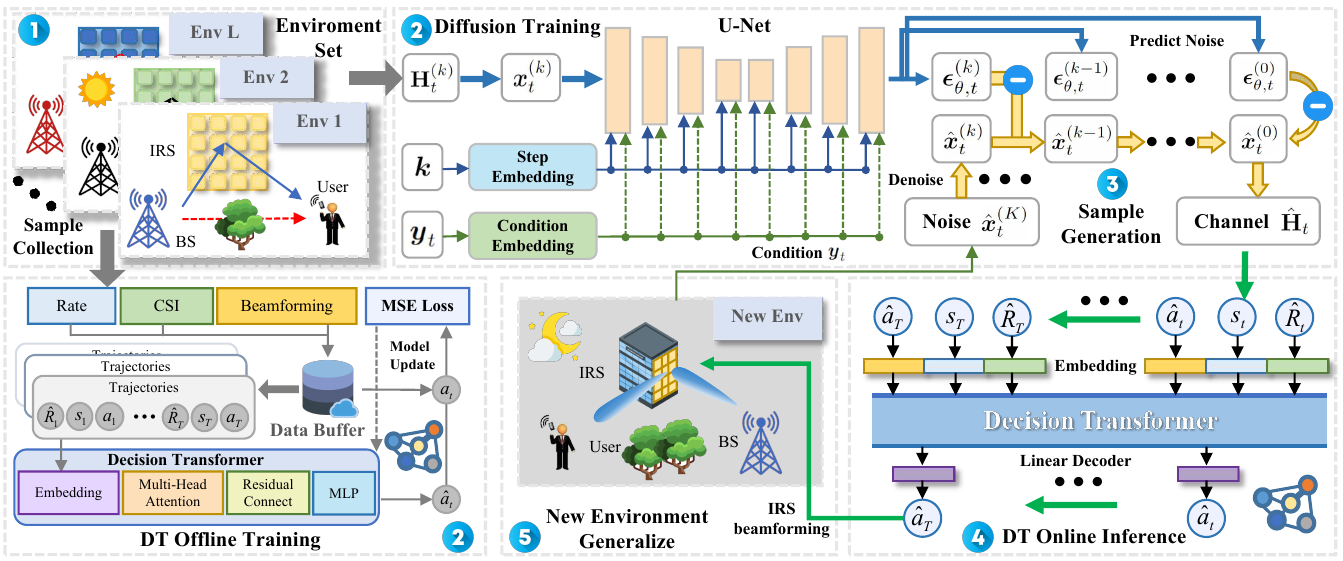}
    \caption{Diffusion-decision transformer structure. It collects data across different IRS environments for offline pre-training of the decision transformer network. The model processes trajectories via a transformer to predict beamforming actions. A U-net-based diffusion model is also trained to obtain channel samples with conditional inputs. The model swiftly adapts to the new environment through fine-tuning to enhance performance.}
    \label{Proposed Architecture}
\end{figure*}

\subsection{Diffusion Model}
DM is a deep generative model designed for the production of high-quality samples. It consists of two distinct processes: the forward diffusion process and the reverse denoising process. During the forward diffusion process, original data is incrementally transformed into an approximated Gaussian noise, while the reverse denoising process reconstructs the original data from the Gaussian noise. In this study, we utilize DM to obtain CSI samples that closely resemble the actual channels. Specifically, the cascaded channel matrix $\boldsymbol{\mathrm{H}}^{(0)}_{t}\in\mathbb{C}^{N\times M}$ (the subscript denotes the diffusion step) can be vectorized, resulting in a real-valued data vector $\boldsymbol{x}^{(0)}_{t}\in\mathbb{R}^{2NM\times 1}$ comprising the real and imaginary parts of the channel matrix. We then define the forward diffusion process as
\begin{equation}
    \boldsymbol{x}^{(k+1)}_{t}=\sqrt{1-\beta_{k}}\boldsymbol{x}^{(k)}_{t}+\sqrt{\beta_{k}}\boldsymbol{\epsilon}^{(k)}_{t}, k=0, \dots, K-1,
\end{equation}
where $\boldsymbol{x}^{(k)}_{t}$ is the noisy data at diffusion step $k$, $\beta_{k}$ is the variance of the added noise at step $k$, and $\boldsymbol{\epsilon}^{(k)}_{t}$ is a Gaussian noise with zero mean and unit variance. As $K$ becomes sufficiently large, $\boldsymbol{x}^{(K)}_{t}$ ultimately follows an approximate Gaussian distribution. If the distribution $q(\boldsymbol{x}_{t}^{(k)}|\boldsymbol{x}_{t}^{(k+1)})$ can be obtained during the denoising process, it is possible to progressively recover the original data. 
We utilize neural networks $q_{\theta}(\boldsymbol{x}_{t}^{(k)}|\boldsymbol{x}_{t}^{(k+1)})$ to predict this intractable distribution~\cite{Gaussian_distribution_for_reverse}. The approximate mean and variance are respectively expressed as

\begin{align}\label{DM mean and variance}
    \mu_{\theta}(\boldsymbol{x}_{t}^{(k)},k)&=\frac{1}{\sqrt{1-\beta_{k}}}\left(\boldsymbol{x}_{t}^{(k)}-\frac{\beta_{k}}{\sqrt{1-\bar{\alpha}_{k}}}\boldsymbol{\epsilon}_{\theta}(\boldsymbol{x}_{t}^{(k)},k)\right),\nonumber\\
    \Sigma_{\theta}(\boldsymbol{x}_{t}^{(k)},k)&=\frac{1-\bar{\alpha}_{k-1}}{1-\bar{\alpha}_{k}}\beta_{k},
\end{align}
where $\bar{\alpha}_{k}=\prod_{i=1}^{k}(1-\beta_{i})$ and $\boldsymbol{\epsilon}_{\theta}(\boldsymbol{x}_{t}^{(k)},k)$ is the predicted noise output by the neural network when it takes the data $\boldsymbol{x}_{t}^{(k)}$ and its corresponding step point $k$ as inputs. 

Our objective is to train the model for accurately predicting channels based on the collected CSI dataset. According to \cite{DDPM}, the corresponding target loss can be simplified to the Kullback-Leibler (KL) divergence between the distribution $q(\boldsymbol{x}_{t}^{(k)}|\boldsymbol{x}_{t}^{(k+1)})$ and the approximate distribution $q_{\theta}(\boldsymbol{x}_{t}^{(k)}|\boldsymbol{x}_{t}^{(k+1)})$ at each diffusion step 
$k$, i.e.,
\begin{equation}\label{Typical DM KL loss}
    L_{k}=\mathrm{D}_{\mathrm{KL}}\left[q(\boldsymbol{x}_{t}^{(k)}|\boldsymbol{x}_{t}^{(k+1)})||q_{\theta}(\boldsymbol{x}_{t}^{(k)}|\boldsymbol{x}_{t}^{(k+1)})\right].
\end{equation}
Finally, the training process is to minimize the gap between the network's predicted noise and the actual added noise, i.e.,
\begin{equation}\label{Typical DM loss}
    L_{k} = \mathbb{E}\left[||\boldsymbol{\epsilon}_{t}^{(k)}-\boldsymbol{\epsilon}_{\theta}(\boldsymbol{x}_{t}^{(k)},k)||^{2}\right].
\end{equation}


\subsection{Decision Transformer}
In this subsection, we firstly map the critical elements of the optimization problem $\boldsymbol{\mathrm{P}}$ into tuples under the framework of RL, including states, actions, and rewards. Specifically, the state $s_{t}$ is defined as the obtained CSI at each moment, the action $a_{t}$ corresponds to the phase shift matrix $\boldsymbol{\Phi}_{t}$, and the reward $r_{t}$ is the data rate received by the user. Traditional RL methods focus on maximizing long-term cumulative rewards through learning a series of actions based on the state. However, the policy learned by this approach often has limited generalizability and tends to be ineffective in new scenarios. In light of this, we shift our focus to the DT, i.e., a model built upon the transformer architecture~\cite{attention_is_all_your_need}, whose core idea is to reframe RL tasks as sequence modeling problems. The DT predicts the optimal future action based on the sequences of historical states, actions, and rewards through an autoregressive model. Specifically, we define the sample sequence (also known as trajectory) as
\begin{equation}
    \tau_{t} = \left(\hat{R}_{1},s_{1},a_{1},\dots,\hat{R}_{t},s_{t},a_{t}\right),
\end{equation}
where $\hat{R}_{t}=\sum_{t'=t}^{T}r_{t'}$ is the cumulative future rewards from time slot $t$ to maximum time slot $T$. It indicates that the model tends to generate actions based on future desired return.

\subsection{Diffusion-Decision Transformer Structure}
Inspired by the DM and DT model, we propose a diffusion-decision transformer (D2T) structure to learn a generalized IRS beamforming policy, as illustrated in Fig.~\ref{Proposed Architecture}.

The structure begins with the collection of environmental samples, where each environment, denoted by $\mathrm{Env}$~$1$, $\mathrm{Env}$~$2$, $\dots$, $\mathrm{Env}$~$L$, differs in aspects such as the distribution of CSI. Data gathered from the historical interactions within these environments encompasses a variety of elements, including IRS beamforming strategies, CSI, and user data rates\footnote{Existing methods such as alternating least squares in ~\cite{Channel_Estimation_method_1,Channel_Estimation_method_2} can be used to collect estimated CSI.}. These pieces of sequential information form trajectories, which are subsequently stored in a data buffer $\mathcal{B}$. It is noted that the collected action samples are either the output of RL or conventional optimization methods that have achieved convergence. This collection phase is critical for capturing the heterogeneity and dynamics of different IRS-aided wireless communication scenarios, ensuring a rich dataset for model training.

Furthermore, the model undergoes offline training by sampling trajectories from the data buffer. Each trajectory $\tau_{t}$, comprising states, actions, and return-to-go, is tokenized and then combined with the position encoding corresponding to the temporal step in the trajectory. These encoded embeddings are fed into a transformer network, wherein the multi-head attention mechanism captures the features and dependencies within the trajectory sequences. The residual connection is also employed to mitigate issues like gradient vanishing and explosion that are common in deep neural networks. The action $\hat{a}_{t}$ is then obtained via a linear decoder composed of multi-layer perceptron (MLP) networks. Finally, the action generation network is optimized by minimizing the mean squared error (MSE) loss between the predicted actions $\hat{a}_{t}$ and their actual counterparts $a_{t}$, i.e.,
\begin{equation}\label{DT loss}
    L(\omega) = \mathbb{E}_{\tau_{t}\in \tilde{\mathcal{B}}}\left[(f_{\omega}(\tau_{t})-a_{t})^{2}\right],
\end{equation}
where $\omega$ is the parameter of the DT model $f_{\omega}$ and $\tilde{\mathcal{B}}$ is a batch sampled from the buffer $\mathcal{B}$. This training phase focuses on learning optimal beamforming matrix $\boldsymbol{\Phi}_{t}$ to maximize the user data rate, given historical sequence of the channel $\boldsymbol{\mathrm{H}}_{1:t-1}$ and the anticipated return-to-go $\hat{R}_{1:t-1}$.

\begin{algorithm}
\caption{The D2T training algorithm}\label{D2T algorithm pseudo code}
\KwIn{Randomly initialize the network parameters $\omega_{0}$ and $\theta_{0}$, the replay buffer $\mathcal{B}$, the diffusion step $K$, the added noise's variance $\beta_{k},\forall k$, the learning rate $\lambda_{1}$, $\lambda_{2}$, the iteration numbers $I_{1}$, $I_{2}$, and the maximum time slots $T$.}
\KwOut{The optimal IRS beamforming matrix $\boldsymbol{\Phi}_{t}$.}
\For{$\mathrm{Env}$ $n=1,2,\dots,L$}{
Interact with the environment with well-trained RL policies to get the reward. Store the trajectory $\tau_{t}$ in the replay buffer $\mathcal{B}$.
}
\For{$\mathrm{each}$ $\mathrm{iteration}$ $i = 1,2,\dots,I_{1}$}{
    Get a batch set $\tilde{B}$ from the replay buffer $\mathcal{B}$.\\
    \For{$t = 1,2,\dots,T$}{
    Get predicted action $\hat{a}_{t}=f_{\omega}(\tau_{t}), \forall \tau_{t}\in\tilde{B}$.\\
    }
    Compute loss $L(\omega) = \mathbb{E}_{\tau_{t}\in \tilde{\mathcal{B}}}\left[(f_{\omega}(\tau_{t})-a_{t})^{2}\right]$.\\
    Update the parameter $\omega_{i+1}\leftarrow \omega_{i}-\lambda_{1}\nabla_{\omega}L(\omega_{i})$.\\
}
\For{$\mathrm{each}$ $\mathrm{iteration}$ $i = 1,2,\dots,I_{2}$}{
    Randomly sample the cascaded channel $\boldsymbol{\mathrm{H}}_{t}$ from the replay buffer $\mathcal{B}$ and calculate the signals $\boldsymbol{y}_{t}$.\\
    Transform the channel $\boldsymbol{\mathrm{H}}_{t}$ into $\boldsymbol{x}_{t}^{0}$.\\
    \For{$k = 0,2,\dots,K-1$}{
        Generate the Gaussian noise $\boldsymbol{\epsilon}_{t}^{(k)}$.\\
        Get the noisy data $\boldsymbol{x}^{(k+1)}_{t}=\sqrt{1-\beta_{k}}\boldsymbol{x}^{(k)}_{t}+\sqrt{\beta_{k}}\boldsymbol{\epsilon}^{(k)}_{t}$.\\
        Compute loss $L^{\mathrm{cond}}_{k}(\theta) = \mathbb{E}\left[||\boldsymbol{\epsilon}_{t}^{(k)}-\tilde{\epsilon}^{(k)}_{\theta,t}||^{2}\right]$.\\
    }
    Update the parameter $\theta_{i+1}\leftarrow \theta_{i}-\lambda_{2}\sum_{k}\nabla_{\theta}L^{\mathrm{cond}}_{k}(\theta_{i})$.
}
\end{algorithm}

Regarding the generation of high-quality channel samples, channel data $\boldsymbol{x}_{t}^{(0)}$ obtained solely from Gaussian noise may possess randomness and cannot accurately estimate the channel. To guide model generation, we utilize the received signals $\boldsymbol{y}_{t}$ generated from pilot signals as auxiliary inputs to the network, which is similar to the conditional inputs in~\cite{IRS_GAN_2}. Furthermore, the distribution, redefined as $q(\boldsymbol{x}_{t}^{(k)}|\boldsymbol{x}_{t}^{(k+1)},\boldsymbol{y}_{t})$ and according to Bayesian formula, can be rewritten as
\begin{equation}\label{condition reverse process 1}
    q(\boldsymbol{x}_{t}^{(k)}|\boldsymbol{x}_{t}^{(k+1)},\boldsymbol{y}_{t})=\frac{q(\boldsymbol{y}_{t}|\boldsymbol{x}_{t}^{(k)},\boldsymbol{x}_{t}^{(k+1)})q(\boldsymbol{x}_{t}^{(k)}|\boldsymbol{x}_{t}^{(k+1)})}{q(\boldsymbol{y}_{t}|\boldsymbol{x}_{t}^{(k+1)})}.
\end{equation}
As $\boldsymbol{x}_{t}^{(k+1)}$ and $\boldsymbol{y}_{t}$ are known at the denoising step $k+1$, the distribution $q(\boldsymbol{y}_{t}|\boldsymbol{x}_{t}^{(k+1)})$ is independent of $\boldsymbol{x}_{t}^{k}$. Hence, (\ref{condition reverse process 1}) can be simplified as 
\begin{equation}
    q(\boldsymbol{x}_{t}^{(k)}|\boldsymbol{x}_{t}^{(k+1)},\boldsymbol{y}_{t})=Cq(\boldsymbol{y}_{t}|\boldsymbol{x}_{t}^{(k)},\boldsymbol{x}_{t}^{(k+1)})q(\boldsymbol{x}_{t}^{(k)}|\boldsymbol{x}_{t}^{(k+1)}),
\end{equation}
where $q(\boldsymbol{y}_{t}|\boldsymbol{x}_{t}^{(k+1)})$ is regarded as a constant $C$. The distribution $q(\boldsymbol{y}_{t}|\boldsymbol{x}_{t}^{(k)},\boldsymbol{x}_{t}^{(k+1)})$ can also be expressed as 
\begin{equation}\label{condition reverse process 2}
    q(\boldsymbol{y}_{t}|\boldsymbol{x}_{t}^{(k)},\boldsymbol{x}_{t}^{(k+1)})=\frac{q(\boldsymbol{x}_{t}^{(k)}|\boldsymbol{y}_{t},\boldsymbol{x}_{t}^{(k+1)})q(\boldsymbol{y}_{t}|\boldsymbol{x}_{t}^{(k+1)})}{q(\boldsymbol{x}_{t}^{(k)}|\boldsymbol{x}_{t}^{(k+1)})}.
\end{equation}
Taking the logarithm and the gradient of (\ref{condition reverse process 2}) yields
\begin{equation}
\begin{aligned}
    \nabla\log{q(\boldsymbol{y}_{t}|\boldsymbol{x}_{t}^{(k)},\boldsymbol{x}_{t}^{(k+1)})}\propto &\nabla\log{q(\boldsymbol{x}^{(k)}_{t}|\boldsymbol{y}_{t},\boldsymbol{x}_{t}^{(k+1)})}\\&-\nabla\log{q(\boldsymbol{x}_{t}^{(k)}|\boldsymbol{x}_{t}^{(k+1)})}.
\end{aligned}
\end{equation}

Similar to (\ref{Typical DM KL loss}), we can approximate the distribution $q(\boldsymbol{x}_{t}^{(k)}|\boldsymbol{x}_{t}^{(k+1)},\boldsymbol{y}_{t})$ with a parameterized network $q_{\theta}(\cdot)$. The corresponding gradient representation can be approximated as
\begin{align}
    &\nabla_{\theta}\log{q_{\theta}(\boldsymbol{x}_{t}^{(k)}|\boldsymbol{x}_{t}^{(k+1)},\boldsymbol{y}_{t})}\approx \eta\nabla_{\theta}\log{q_{\theta}(\boldsymbol{x}_{t}^{(k)}|\boldsymbol{y}_{t},\boldsymbol{x}_{t}^{(k+1)})}\nonumber\\&\qquad\qquad\qquad+(1-\eta)\nabla_{\theta}\log{q_{\theta}(\boldsymbol{x}_{t}^{(k)}|\boldsymbol{x}_{t}^{(k+1)})},
\end{align}
where $\eta$ is a coefficient measuring the importance of the conditional inputs $\boldsymbol{y}_{t}$. Consequently, we modify the loss function $L_{k}$ as
\begin{equation}\label{Modified DM Loss}
    L^{\mathrm{cond}}_{k}(\theta) = \mathbb{E}\left[||\boldsymbol{\epsilon}_{t}^{(k)}-\tilde{\epsilon}^{(k)}_{\theta,t}||^{2}\right].
\end{equation}
In~(\ref{Modified DM Loss}), $\tilde{\epsilon}^{(k)}_{\theta,t}$ is the modified predicted noise, i.e.,
\begin{equation}
    \tilde{\epsilon}^{(k)}_{\theta,t}=\eta\boldsymbol{\epsilon}_{\theta}(\boldsymbol{x}_{t}^{(k)},k,\boldsymbol{y}_{t})+(1-\eta)\boldsymbol{\epsilon}_{\theta}(\boldsymbol{x}_{t}^{(k)},k,\emptyset),
\end{equation}
where $\boldsymbol{\epsilon}_{\theta}(\boldsymbol{x}_{t}^{(k)},k,\boldsymbol{y}_{t})$ represents the noise predicted with the additional conditional inputs $\boldsymbol{y_t}$. Let us encode the diffusion step $k$ and the received signals $\boldsymbol{y}_{t}$ by means of the step embedding and condition embedding modules, respectively. These encoded inputs are then combined with the data $\boldsymbol{x}_{t}^{(k)}$ and fed into the U-Net network. To effectively integrate and understand the features of these three types of inputs, we employ one-dimensional convolutional networks. Finally, the channel generation network is updated by minimizing (\ref{Modified DM Loss}). 

In the new environment for inference, we first obtain the generated channel samples. It begins with a Gaussian noise $\boldsymbol{x}_{t}^{(K)}$ and progressively removes the predicted noise $\tilde{\boldsymbol{\epsilon}}_{\theta,t}^{(k)}$. To ensure the stability and quality of the generation process, we utilize the reparameterization trick to generate the data, i.e.,
\begin{equation}\label{reparameterization generation data}
    \boldsymbol{x}_{t}^{(k)}=\frac{1}{\sqrt{1-\beta_{k}}}\left(\boldsymbol{x}_{t}^{(k+1)}-\frac{\beta_{k}}{\sqrt{1-\bar{\alpha}_{k}}}\tilde{\boldsymbol{\epsilon}}_{\theta,t}^{(k)}\right)+\Sigma_{\theta}(\boldsymbol{x}_{t}^{(k)},k)z_{t},
\end{equation}
where $z_{t}$ is a Gaussian noise. Through $K$ rounds of denoising, the channel state $s_{t}$ is obtained. Furthermore, we utilize a small set of samples generated by suboptimal strategies, such as RL models that are only partially trained, to fine-tune the pre-trained D2T model. Then the D2T model can iteratively get the action $\hat{a}_{t}$, which are informed by real-time rate feedback $r_{t}$, updated channel information $s_{t}$, and historical trajectory $\tau_{t-1}$, leading to a series of IRS beamforming strategies that improve communication performance. Using the pre-trained knowledge and the generative capabilities to adjust strategies, the proposed method is a generalized adaptation process tailored to the dynamic characteristics of the new environment. 

\begin{algorithm}
\caption{The D2T inference algorithm}\label{D2T algorithm inference pseudo code}
\KwIn{New $\mathrm{Env}$, few-shot samples $\hat{B}$, diffusion step $K$, added noise's variance $\beta_{k},\forall k$, maximum time slots $T$, target return-to-go $\hat{R}_{1}$, and initial input sequence $\tau^{\mathrm{in}}_{1}=(\hat{R}_{1})$.}
\KwOut{The optimal IRS beamforming matrix $\boldsymbol{\Phi}_{t}$.}
Fine-tune the pre-trained D2T model with $\hat{B}$.\\
\For{$t=1,2,\dots, T$}{
    Get the signal $\boldsymbol{y}_{t}$ and random Gaussian noise $\boldsymbol{x}_{t}^{(K)}$.\\
    \For{$k=K-1,\dots,0$}{
    Predict noise $\tilde{\boldsymbol{\epsilon}}_{\theta,t}^{(k)}$ and get noise $z_{t}$.\\
    Denoise the data following~(\ref{reparameterization generation data})}
    Transform the data $\boldsymbol{x}_{t}^{0}$ into channel state $s_{t}$.\\
    Update the sequence $\tau^{\mathrm{in}}_{t}\leftarrow(\tau^{\mathrm{in}}_{t},s_{t})$.\\
    Get predicted action $\hat{a}_{t}=f_{\omega}(\tau^{\mathrm{in}}_{t})$.\\
    Interact with the environment to get reward $r_{t}$.\\
    Update the return-to-go $\hat{R}_{t+1}=\hat{R}_{t}-r_{t}$.\\
    Update the sequence $\tau^{\mathrm{in}}_{t+1}\leftarrow (\tau^{\mathrm{in}}_{t},\hat{a}_{t},\hat{R}_{t+1}).$
}
\end{algorithm}

The training process of the proposed D2T method is summarized in Algorithm~\ref{D2T algorithm pseudo code}. Firstly, we initialize the neural network parameters and the hyper-parameters. D2T interacts with multiple environments to collect trajectory samples, which are then stored in an experience replay buffer (see lines $1$-$2$). Subsequently, samples from this buffer are used to train the D2T (see lines $3$-$16$). The inference process for new environments is described in Algorithm~\ref{D2T algorithm inference pseudo code}. Specifically, we begin with setting a target return-to-go and then obtain channel data using the trained model in each time slot based on the given condition signals. After fine-tuning the pre-trained D2T model, we further obtain optimal phase shift strategies in a new environment.

\section{Simulation Results}
In this section, simulation results are used to evaluate the performance of the proposed D2T framework. The default simulation parameters are set as follows. Let $\beta_{t,n}=1$. The channels between the BS and IRS, as well as that between the IRS and the user, are both assumed to be Rician Fading, i.e.,
\begin{equation}\label{Ricial Channel}
\begin{aligned}
    \boldsymbol{\mathrm{G}}_{t} &= \sqrt{\mathcal{L}_{\mathrm{1}}}\left(\sqrt{\frac{\kappa_{1}}{1+\kappa_{1}}}\bar{\boldsymbol{\mathrm{G}}}_{t} + \sqrt{\frac{1}{1+\kappa_{1}}}\Tilde{\boldsymbol{\mathrm{G}}}_{t}\right),\\
    \boldsymbol{\mathrm{h}}_{t} &= \sqrt{\mathcal{L}_{\mathrm{2}}}\left(\sqrt{\frac{\kappa_{2}}{1+\kappa_{2}}}\bar{\boldsymbol{\mathrm{h}}}_{t} + \sqrt{\frac{1}{1+\kappa_{2}}}\Tilde{\boldsymbol{\mathrm{h}}}_{t}\right),
\end{aligned}
\end{equation}
where $\kappa_{1}$ and $\kappa_{2}$ are the Rician factors, $\bar{\boldsymbol{\mathrm{G}}}_{t}$ and $\bar{\boldsymbol{\mathrm{h}}}_{t}$ denote the deterministic line-of-sight (LOS) components, and $\Tilde{\boldsymbol{\mathrm{G}}}_{t}$ and $\Tilde{\boldsymbol{\mathrm{h}}}_{t}$ are the Rayleigh fading. The path loss models for the BS-IRS link and the IRS-user link are expressed as $\mathcal{L}_{1}=\mathcal{L}_{0}-10\xi_{1}\log_{10}{(d_{1}/d_{0})}$ dB and $\mathcal{L}_{2}=\mathcal{L}_{0}-10\xi_{2}\log_{10}{(d_{2}/d_{0})}$ dB, respectively, where  $\mathcal{L}_{0}=-30$ dB is the path loss at the reference distance $d_{0}$, $\xi_{1}, \xi_{2}$ are the path loss exponents, and $d_{1}, d_{2}$ are the BS-IRS and IRS-user distances. The BS is equipped with a uniform linear array antenna with $M=4$. The noise power and transmit power are respectively set to $n_{t}=-90$ dBm and $P=5$ dBm. The channel factors are set to $\kappa_{1}=\kappa_{2}=10$, $\xi_{1}=2.2$, and $\xi_{2}=2.8$. We utilize the default parameter settings as the simulation environment for offline pre-training. Furthermore, we change the channel parameters and BS-IRS distances to simulate different wireless environments, such as urban and suburban areas.
\begin{figure}[htbp]
	\vspace{-0.2cm}
   \centering
   \includegraphics[width=0.48\textwidth]{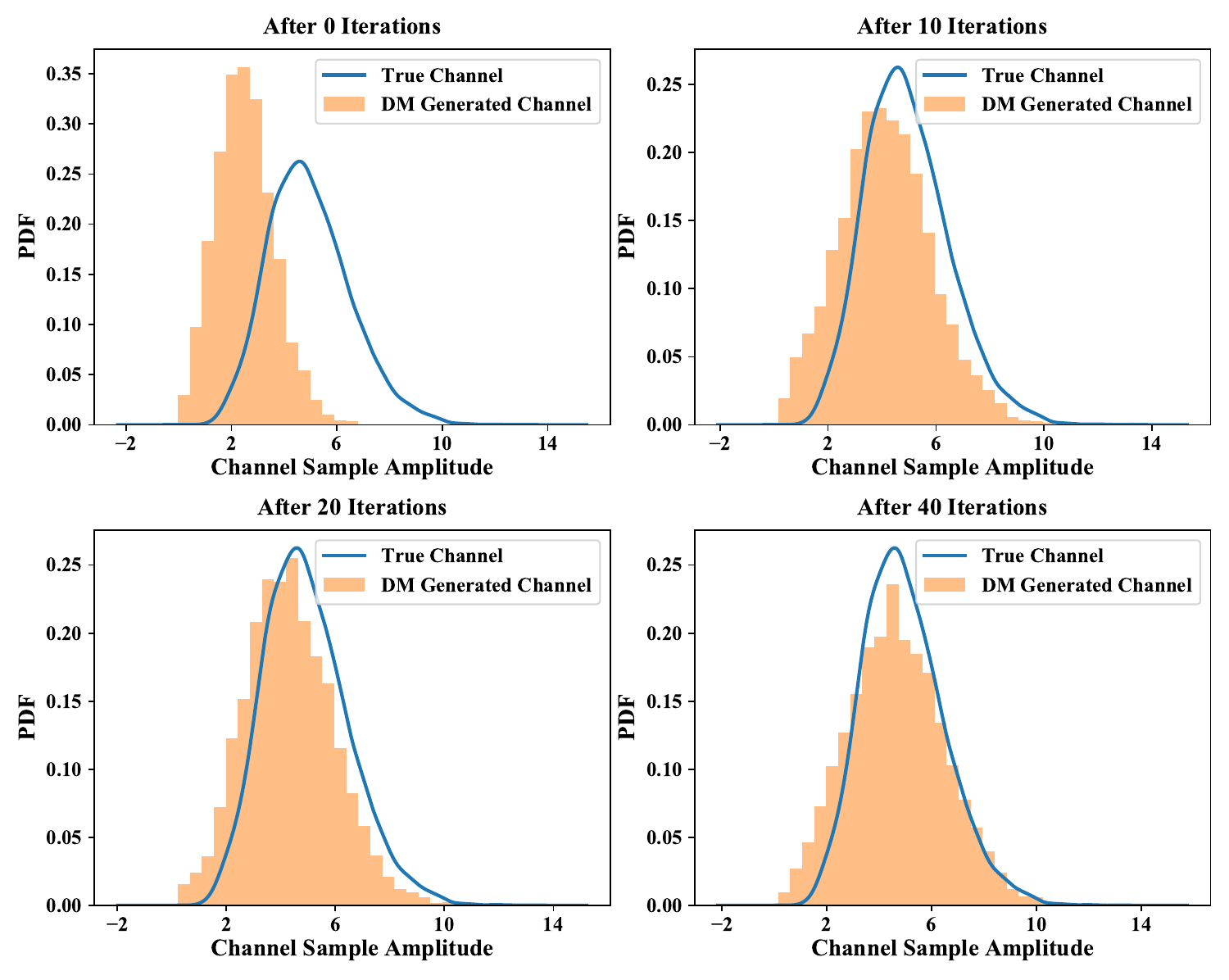}
   \caption{PDF comparisons of the true channel samples and the generated channel samples by DM during the training process.} 
   \label{PDF comparison}
\end{figure}

Then, we introduce the detailed network settings of the proposed D2T. For the DM part, we conducted diffusion training for $K=500$ steps, with the noise parameter $\beta_{k}$ ranging from $10^{-4}$ to $0.02$. The hidden layer dimension of the time embedding and conditional embedding modules is set to $512$. The U-Net network consists of $6$ downsample blocks for encoding input feature, along with $6$ upsample blocks for decoding. Each sampling module is composed of one-dimensional convolutional layers and linear layers. The DT comprises $3$ transformer blocks with a hidden layer dimension of $256$ and a dropout rate of $0.1$. Our model is trained on NVIDIA RTX A6000 GPU hardware, employing the AdamW optimizer with a learning rate of $10^{-4}$.

We illustrate the channel generation process of our proposed method in Fig.~\ref{PDF comparison}, comparing the probability density function (PDF) of the model's predicted channel with that of the true ones. At the beginning of the training process, the samples generated by the DM are significantly different from the true samples, which is due to the random initialization of the network. Subsequently, the proposed network gradually learns to approximate the true channel. After $40$ iterations of training, the obtained channels closely coincide with the real channel samples. This observation indicates that the proposed method has successfully predicted the channel.



\begin{figure}[htbp]
   \centering
   \vspace{-0.3cm}
   \includegraphics[width=0.48\textwidth]{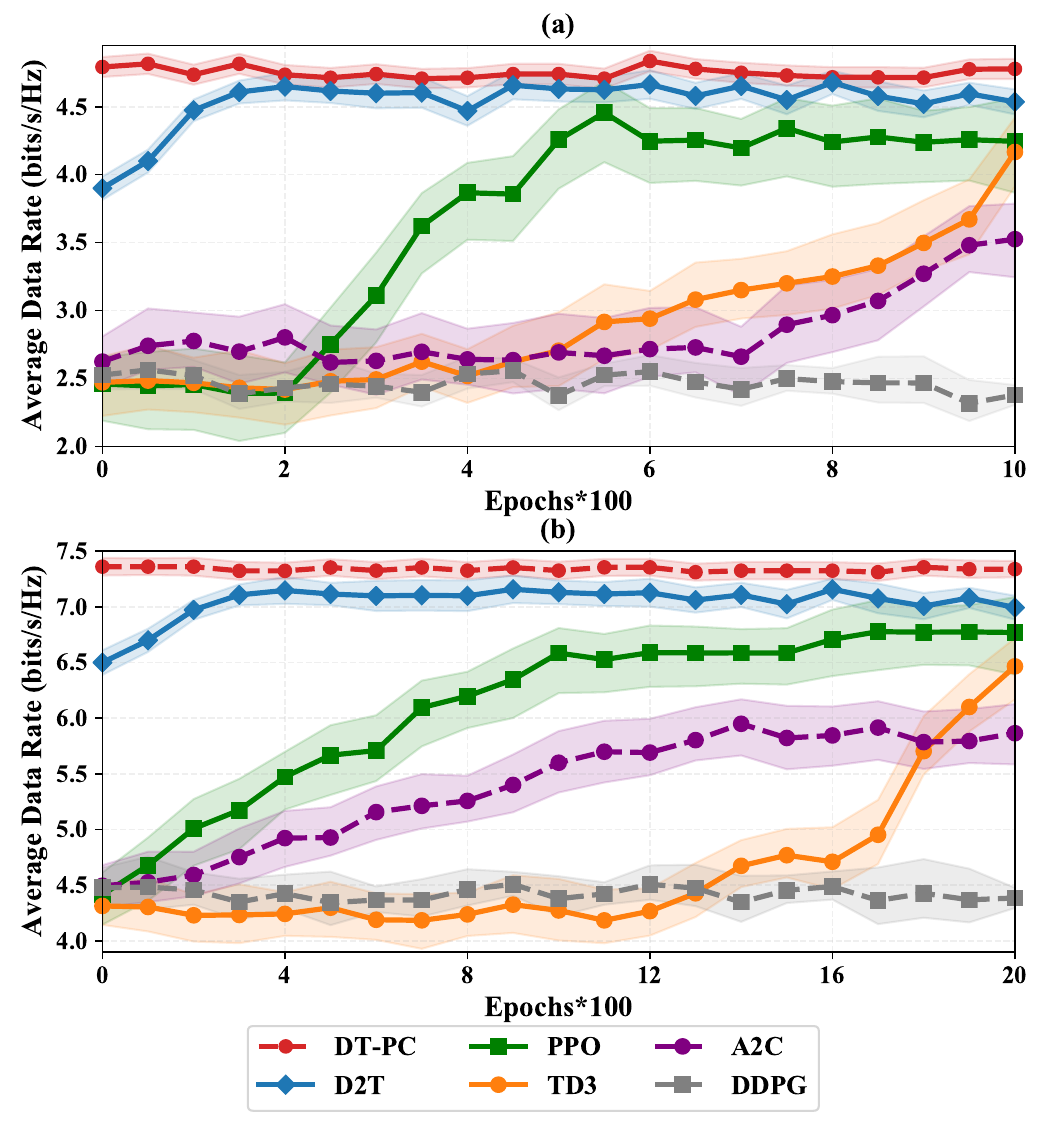}
   \caption{Performance and convergence speed comparison between D2T and traditional RL methods. (a) IRS element $N=64$. (b) IRS element $N=128$.} 
   \label{Performance Comparison}
\end{figure}

In Fig.~\ref{Performance Comparison}, we plot the average achievable rate performance of the proposed D2T in the new environment with various IRS elements. For comparison, we provide the performance achieved by the typical RL methods~\cite{RL_survey}, i.e., proximal policy optimization (PPO), twin delayed DDPG (TD3), advantage actor critic (A2C), and DDPG. It is observed that D2T initially performs well even with zero-shot learning. It achieves $6\%$ performance improvement as well as $3$ times faster in terms of convergence speed than PPO. RL methods necessitate retraining with a slower convergence rate when confronted with a new environment. This is because the D2T does not rely on iterative approximations of value functions, it directly learns the complex mappings between trajectory sequences and optimal actions. The fine-tuned DT trained with perfect CSI, labeled as DT-PC, serves as an upper bound under the structure of DT. It is observed that D2T is very close to this bound, which substantiates the capability of D2T to output channels that closely approximate the real ones. Through pre-training across various environments, D2T can generate policies that exhibit superior generalization capabilities. Consequently, superior to conventional RL methods, D2T can rapidly adapt to new environments without training from scratch.

\section{Conclusions}
In this study, we have proposed the innovative D2T method to enhance IRS beamforming optimization across varying channel distributions environments. The D2T method employs the DM that uses real-time received signals from pilot data to accurately obtain the realistic channel states. This approach moves away from traditional channel estimation methods that rely on prior theoretical models, and thereby is more efficient in channel acquisition. The generalized decision model is developed through a pre-training process that utilizes datasets collected from RL optimization across diverse channel environments. In new channel distribution scenarios, our method effectively generalizes strategies with a limited number of samples for fine-tuning. Simulation results corroborate that the D2T algorithm outperforms traditional RL methods in terms of efficiency, presenting a fast adaptation to new channel distributions.

\bibliographystyle{IEEEtran}
\bibliography{IEEEabrv,Ref}

\end{document}